\documentclass[aps,prl,twocolumn,superscriptaddress,floatfix]{revtex4}
\usepackage{amssymb}
\usepackage{graphicx}
\usepackage[english]{babel}
\usepackage{latexsym}
\usepackage{graphics}
\usepackage{subfigure}

\def\be{\begin{equation}}
\def\ee{\end{equation}}
\def\bea{\begin{eqnarray}}
\def\eea{\end{eqnarray}}
\def\bse{\begin{subequations}}
\def\ese{\end{subequations}}

\def\be{\begin{eqnarray}}
\def\ee{\end{eqnarray}}

\begin{document}
\title{Ultra-high fidelity qubits for quantum computing}
\author{Mark G. Raizen} \email{raizen@physics.utexas.edu}
\author{Shou-Pu Wan}
\affiliation{Center for Nonlinear Dynamics, The University of
Texas at Austin, Austin, Texas 78712 USA} \affiliation{Department
of Physics, The University of Texas at Austin, Austin, Texas 78712
USA}
\author{Chuanwei Zhang} \affiliation{Department of Physics and
Astronomy, Washington State University, Pullman, Washington, 99164
USA}

\author{Qian Niu}\affiliation{Department of Physics, The
University of Texas at Austin, Austin, Texas 78712 USA}

\begin{abstract}
We analyze a system of fermionic $^{6}$Li atoms in an optical
trap, and show that an atom ``on demand'' can be prepared with
ultra-high fidelity, exceeding $0.99998$. This process can be
scaled to many sites in parallel, providing a realistic method to
initialize N qubits at ultra-high fidelity for quantum computing.
We also show how efficient quantum gate operation can be
implemented in this system, and how spatially resolved single-atom
detection can be performed.
\end{abstract}
 \maketitle

The potential impact of quantum computing has stimulated a
worldwide effort to develop the necessary experimental and
theoretical resources \cite{book:qc:device,book:qc:theory}. The
most challenging aspect of quantum computing is the requirement
for ultra-high fidelity in each step. The necessary starting point
is initialization of a scalable number of qubits at ultra-high
fidelity. In this Letter, we consider a system of fermionic atoms
and show that an atom ``on demand'' can be prepared in the ground
state of a trap with fidelity exceeding $0.99998$, providing a
realistic solution to the initialization problem in quantum
computing.  We discuss how that process can be scaled to many
sites in parallel, enabling pairwise entanglement operations.
Finally we show how spatially resolved detection of each state can
be implemented.

We first address the question of how to trap one atom ``on
demand'' in the ground state of a trap. Note that such a process
should be scalable up to many optical traps in parallel. A single
atom``on demand'' in the ground state has not yet been
demonstrated experimentally. However, progress has been made for
bosonic atoms using the method of laser culling \cite{Dudarev,
chuu:2005,pons:2009}. One of the key questions is that of fidelity
of the number state, and in that regard bosons are not ideal
because they rely on strong interactions to maintain a relatively
large excitation gap, and to suppress low-frequency excitations
during the culling process. This leads us to propose instead
fermionic atoms where a precise number is rigorously enforced by
the Pauli exclusion principle. More specifically, we propose to
use $^{6}$Li as the atom of choice. This atom has the advantage
that the interaction strength and sign (attractive or repulsive)
can be tuned with an external magnetic field. Two magnetic
sublevels of one hyperfine ground state, $\left\vert
F=\frac{1}{2},m_{F}=\frac{1}{2 }\right\rangle $ and $\left\vert
F=\frac{1}{2}, m_{F}=-\frac{1}{2} \right \rangle $ are used to
define a qubit. We denote these states as $\left\vert \uparrow
\right\rangle $ and $\left\vert \downarrow \right\rangle $
respectively. These states both become high-field seekers at large
magnetic field with a well-defined frequency splitting that is
nearly field-independent, hence insensitive to magnetic noise.

The starting point of the ``on demand'' single atom preparation is
laser-cooled $^{6}$Li atoms that are optically trapped in the two
spin states \cite{thomas:2000,hulet:1996}. The atoms can be cooled
by evaporation at a magnetic field around $300$ Gauss, where the
scattering length $a_s \approx -300 a_{0}$ ($a_{0}$ is the Bohr
radius). $a_s$ is large enough for efficient evaporation of the
spin mixture, and is also at a minimum as a function of magnetic
field, greatly reducing the effects of magnetic noise. After
evaporative cooling, a weakly interacting degenerate Fermi gas
forms at temperature $T \ll T_{F}$, where $T_{F}$ is the Fermi
temperature. The single atom preparation process can be split into
three steps:
\begin{description}
\item[Step I] The magnetic field is tuned to near $\sim 0$ Gauss
from the initial field of $300$ Gauss, resulting in a
non-interacting degenerate Fermi gas (DFG). In this state, a spin
pair fills each level, up to approximately the Fermi level.
\item[Step II] Atom pairs are ejected by laser culling. This is
accomplished by adiabatic lowering of the optical potential. This
prepares a single pair in the ground state. \item[Step III] The
well is adiabatically split into two parts that are spatially
separated. In the presence of a magnetic field bias this prepares
one spin state on the left and the other on the right. Each atom
can then serve as the initial state for a qubit.
\end{description}
With this method, an array of $N$ micro-traps would prepare $2N$
qubits. We now show in detail how ultra-high fidelity is enforced
at each step.

\begin{figure*}[tbp]
\includegraphics[scale=0.7]{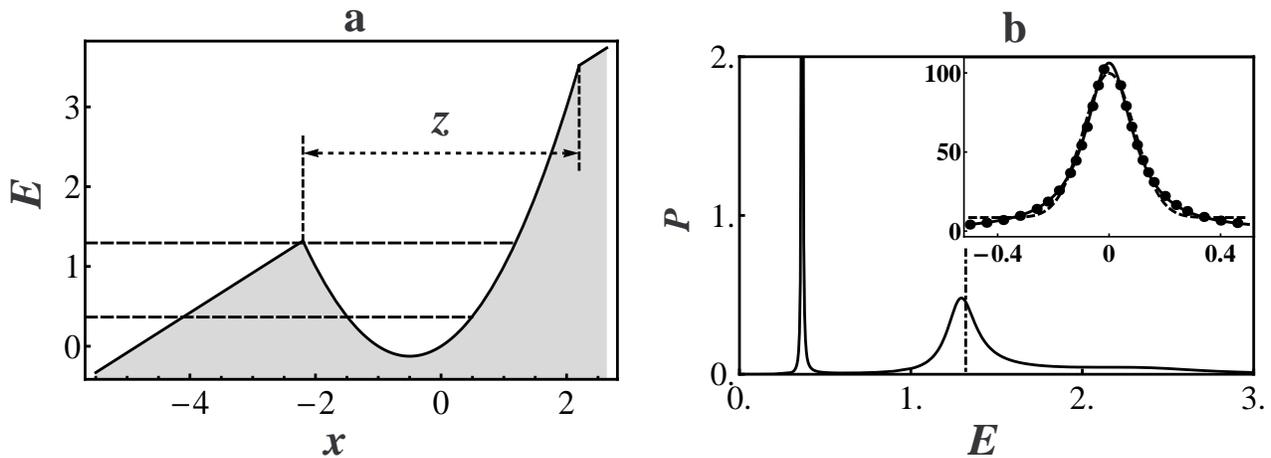}\\
\caption{The simulation model for laser culling in a 1D optical
trap. We take $x_0 \equiv \sqrt{\hbar/m \omega}$ as the unit of
length ($m$ is the mass of the $^6$Li atom), $\hbar\omega$ as the
unit of energy, and $\hbar\omega / x_0$ as the unit of force.
\textbf{a} Truncated harmonic trap with magnetic gradient. The
trapping potential for one state is shown. Parameters used to
specify the trap: trap size, $z$, is the length of the parabolic
portion in the potential profile; force, $f$ is the slope of the
linear portion in the potential profile. Dashed lines denote the
levels of the ground and first-excited states. \textbf{b} Plot of
$P(E)$ as a measure of density of states of the quasi-bound
states. Trap size $z=4.4$, force $f = 0.5$. The vertical line
(dotdashed) denotes the barrier height of the trap. Inset:
Zoomed-in view of the resonance peak of the ground quasi-bound
state (dots) as shown in the main figure and Lorentzian (solid)
and Gaussian (dashed) regressions. Horizontal axis is in units of
$\gamma /2$, where $\gamma$ is the FWHM of the resonance peak. }
\label{fig:DOS}
\end{figure*}

\textbf{Step I} In the presence of a scattering length $a_s
\protect \approx\protect-300a_{0}$ between the $\left\vert
\uparrow \right\rangle $ and $\left\vert \downarrow \right\rangle
$ states, the Fermi gas is weakly interacting. At $T=0$, the Fermi
gas may form a BCS state. The pairing gap for such a state can be
estimated through $\Delta \approx 0.5E_{F}\exp \left( \pi
/2k_{F}a\right)$ \cite{Giorgini:2008}. For a degenerate $^{6}$Li
gas with $k_{F}^{-1}\sim 1000a_{0}$, the pairing gap $\Delta
\approx 0.002E_{F}$. With such a small $ \Delta $, the occupation
probability of the lowest energy state (\textit{e.g.} the
$\mathbf{k=0}$ state for a uniform gas) $n_{\mathbf{k=0}}=\left(
1-\eta _{\mathbf{k}}/\epsilon _{\mathbf{k}}\right) /2\approx
1-4\times 10^{-6}$, where $\eta _{\mathbf{k}}=\varepsilon
_{\mathbf{k}}-E_{F}$, $ \varepsilon _{\mathbf{k}}=\hbar
^{2}k^{2}/2m$, $\epsilon _{\mathbf{k}}=\left( \Delta ^{2}+\eta
_{\mathbf{k}}^{2}\right) ^{1/2}$ is the quasi-particle excitation
energy, and we take the chemical potential $\mu \approx E_{F}$ for
the BCS state. In addition, for such a small pairing gap, finite
temperature effects dominate, and there may even be no BCS
pairing. However, the finite temperature does not affect the
ground state occupation probability. Consider a temperature
$T=0.05T_{F}$, for which the ground state occupation probability
is approximately $1/\left[ \exp \left( -E_{F}/k_{B}T\right)
+1\right] =1-4\times 10^{-5}$. Therefore, a fast sweep of the
magnetic field (i.e., the scattering length) to the
non-interacting region does not affect the ground state occupation
probability or the fidelity of the single atom preparation.

\textbf{Step II} A non-interacting DFG in a deep optical trap
serves as the starting point for the laser culling process. We
assume that the trapping frequencies in the two transverse
directions are significantly higher than along the axis, and only
the ground transverse state can be supported. We therefore limit
our analysis to one spatial dimension. The trap wall is reduced to
a level slightly below the ionization threshold of the
first-excited state of the optical dipole trap. The trap reduction
rate is chosen to fulfill the adiabaticity requirement. To that
end, we maintain a constant trapping frequency, $\omega$,
throughout the laser culling process, which can be accomplished by
dynamically varying the focus of the beam \cite{weiss:2005}. (In
practice, this step may not be necessary, but it greatly
simplifies the calculation.) According to the
Wentzel-Kramers-Brillouin (WKB) method, atoms with energy much
lower than the trap depth are completely unaffected by the change
in the depth of the trap. The adiabaticity condition is fulfilled
as long as the WKB approximation is valid, which holds until the
trap depth is around $3\hbar \omega/2$, where $\hbar \equiv
h/2\pi$ and $h$ is Planck's constant. Near and beyond this point,
the trap reduction rate must be slowed down to maintain
adiabaticity.

A constant force (or tilt), $f$, is also needed to sweep the atoms
away from the micro-traps as soon as they are ionized, which is
generated by applying a magnetic field gradient. For simplicity,
we assume the force has a positive sign for state $\left\vert
\uparrow \right\rangle $. In our simulation, we approximate the
optical trap with a truncated harmonic potential (see
Fig.\ref{fig:DOS}\textbf{a}) which is specified by the \emph{trap
size}, $z $. After reaching a minimum trap size $z_m$, the trap is
held for a certain time to allow ionized atoms to escape, while
keeping the ground-state atom pair with high probability. The trap
depth, magnetic force, and holding time are adjusted to optimize
fidelity. Finally the trap wall (and size) is adiabatically raised
to a higher level to preserve the resultant Fock state.

Our simulation method is described below. We assume that the
ground state fidelity is unchanged until the ionization energy of
the first excited state is approached. Because of the tilt, the
trap has no stationary bound state, only quasi-bound states,
which, as $z \to \infty $, become the bound states. The lifetimes
of the quasi-bound states determine the rate of the change of trap
occupation probability. The optimized final trap depth, tilt, and
holding time is calculated.

Suppose a stream of incoming atoms is incident from $x = -\infty$,
with energy $E$, scattered by the trap potential. Let $\psi _E
\left( x\right) $ be the wavefunction of the stationary state of
the incoming atoms, which is assumed to have unit amplitude at $x
= -\infty$. We assume the trap is located at $[-z/2,z/2]$. Outside
the trap ($x<-z/2$), $\psi_E \left( x\right) $ is a superposition
of Airy functions $\text{Ai}(x)$ and $\text{Bi}(x)$ with necessary
superposition amplitudes and phases; inside the trap,
\begin{equation}
\psi_E \left( x \right) = a(E) e^{-x^2/2} H_\nu (x),
\end{equation}
where $H_\nu(x)$ is the Hermite function of degree $\nu$, $a(E)$
is the amplitude. We require $\psi_E \left( x\right) $ be
continuous and differentiable for all $x$, by which $a(E)$ is
obtained. Of special interest to us are those states that have
significant amplitude in the trap area, because they correspond to
the quasi-bound states of the trap. We define $P(E) \equiv \left
\vert a(E) \right \vert ^2$. For simplicity, we can take $P(E)$ as
a measure of the density-of-states of the trap. We plot $P(E)$ in
Fig. \ref{fig:DOS}\textbf{b}. Note that in the limit of zero
magnetic force, the peaks at $E \approx 0.366 \text{ and } 1.29$
correspond to the ground and first-excited states of the truncated
harmonic trap, respectively.

$P(E)$ describes not only the ionization thresholds, but also the
dynamical properties of the quasi-bound states. To see that, we
study the evolution of a wavefunction $\phi(x,t)$. Imagine that at
time $t=0$, we have
\begin{equation}\label{eq:initWF}
\phi(x,0) = \left\{ \begin{array}{cc}
  c\; \psi_{E_0}(x), & -z/2 < x < z/2; \\
  0, & \text{otherwise,} \\
\end{array}\right.
\end{equation} where $E_0$ is one of the resonance energies, $c$
is a normalization factor such that $\int_{ -\infty}^{ \infty}{
\vert \phi( x,0) \vert ^2 dx} = 1$ and there is no other atom
source. Subsequently, the atom will start to tunnel out of the
trap. The probability, $R_{E_0}(t)$, for an atom to remain in the
trap is given by $\int_{ -z/ 2}^{ z/2} {dx \left \vert \phi(x,t)
\right\vert^2}$. To evaluate $R_{E_0}(t)$, we expand $\phi(x,t)$
in terms of the wavefunctions of the stationary states,
$\psi_E(x)$. Note that in the vicinity of a resonance peak at $E =
E_0$, one can write the wavefunction $\psi_{E}(x) \approx C(E)
a(E) \psi_{E_0}(x)$, where $C(E)$ is a slowly varying quantity.
Also
\begin{equation}
\int_{- z/2}^{ z/2} {dx \psi_ {E}^{*}(x) \psi_{E_0}(x)} \approx
\int_{-\infty}^{\infty}{dx \psi_{E}^{*}(x) \psi_{E_0}(x)}
\end{equation} for $ E \neq E_0$, due to the oscillatory nature of Airy functions.
We find
\begin{equation}\label{eq:RP:2}
R_{E_0}(t) \approx |C(E_0)|^2 \left\vert \int_{E_0 -
\epsilon/2}^{E_0 + \epsilon/2}{ P(E) e^{i E t}  dE}\right\vert^2,
\end{equation}
where $\epsilon$ is the range of integration. In the vicinity of a
resonance peak, the function $P(E)$ is essentially Lorentzian (see
inset of Fig. \ref{fig:DOS}b), and we finally obtain
\begin{equation} \label{eq:RP:3}
R_{E_0}(t) \approx e^{-\gamma_{E_0} t},
\end{equation} where $\gamma_{E_0} $ is the full width at half maximum
(FWHM) of the resonance peak at $E=E_0$. The lifetime of the
quasi-bound state at $E=E_0$ is $\tau_{E_0} =\gamma_{E_0} ^{-1}$.
Eq.(\ref{eq:RP:3}) is used to determine an optimized combination
of minimum trap depth, magnetic force, and holding time for the
best fidelity. It is also worth noting that ultimately the
\emph{difference} in the lifetimes between the ground and the
first excited quasi-bound states determines the fidelity of
producing a pair of atoms in the ground state of the trap.

\begin{figure}[t]
\includegraphics[scale=0.6]{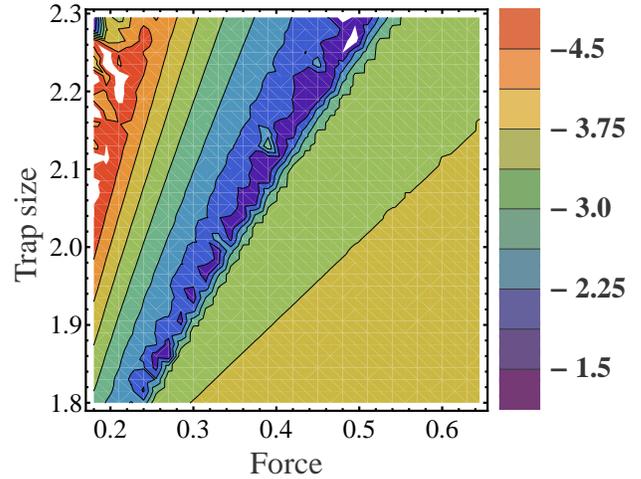}
\caption{Ground state fidelity after laser culling. A varying
holding time is allowed to ensure that the residual probability
for atoms in excited quasi-bound states is no larger than
$10^{-5}$. Shown in the panel is the base-10 logarithm of the
ground state fidelity loss. Red (blue) color represents high (low)
fidelity. The white areas are out-of-range clippings: near the
left-hand side, the white areas should be redder than its
surrounding color; near center-top, the white area should be bluer
than its surrounding color. \label{fig:gnd:fidelity}}
\end{figure}

Taking into account all the steps in laser culling, we show the
fidelity of preparing a single pair of atoms in the ground state
of a micro-trap in Fig \ref{fig:gnd:fidelity}. Choosing culling
parameters near the upper-left corner of the figure results in
ultra-high fidelity. For a set of realistic parameters: trapping
frequency $\omega = 2\pi \times 1$ kHz, magnetic field gradient
$0.66$ Gauss/cm, and truncated trap size $8.8\; \mu$m, the
ground-state to first-excited state lifetime ratio is $7.53\times
10^{5}$. With a holding time of $218$ ms, we get a residual
probability of 10$^{-5}$ for the excited-state and a ground state
occupation probability larger than $0.99998$.

\textbf{Step III} Finally, the pair of atoms are split and trapped
in the ground state of two adjacent micro-traps. In order to
achieve deterministic splitting, where one spin state is driven to
the left and the other to the right, we impose a magnetic field
gradient (providing a force $f$) while we adiabatically split the
optical tweezer (that holds the pair) into two beams
\cite{beugnon:2007}. With an appropriate magnetic field gradient
and a separating barrier, this step can be realized at ultra-high
fidelity. Since $\left\vert \downarrow \right\rangle $ is a
low-field-seeker and $\left\vert \uparrow\right\rangle $ is a
high-field-seeker at low magnetic field, each atom is displaced to
a different location as soon as the trap is split.

We adopt a simpler potential than that of a realistic optical
tweezer. The trap is composed of two spliced sections of parabolic
trap with identical trapping frequency $\omega$, each of which has
an energy minimum, located at $\pm d/2$, respectively, where $d$
is the separation distance. We numerically calculate the
eigen-energy levels and wavefunctions of such a double-well
structure in the parameter space of $f$ and $d$. One needs to
choose a path in the parameter space such that a sufficient gap
between the ground and first-excited states is maintained
throughout the splitting process in order to suppress transitions
from the ground state. With a trapping frequency $\omega =
2\pi\times1$ kHz, a magnetic field gradient of $0.66$ Gauss/cm and
a separation displacement $d = 6.25\; \mu$m, we find $0.99998$
splitting fidelity.

So far, we have shown that two initialized fiducial states can be
prepared at ultra-high fidelity. How can one scale this process to
multiple sites, and make the ``switchyard'' of multiplexed beams
to perform the required complex operations? The scalability of
this system can be achieved using scalable microelectromechanical
systems (MEMS) technology \cite{kim:2009}. Using this technology,
an array of beams can be directed to each site where individual
atoms can be trapped. An alternative approach was developed in
ref. \cite{Dumke:2002}. By steering the beams, one can transport
individual atoms and bring them into pairwise interaction with
arbitrary control. An array of micro-traps would most easily be
accomplished with red-detuned beams that create attractive
potentials along the axis. This technique also enables the
creation of a two-dimensional optical trap array, and entanglement
of any pair using the qubit transfer technique of ref.
\cite{beugnon:2007}. This could extend the linear case where only
nearest-neighbor operations are possible.

We now consider single-qubit gates and two-qubit gates in this
system. The implementation of a single-qubit gate requires the
capability to address each atom individually. This can be
accomplished with stimulated Raman transitions, as is currently
employed with trapped ions \cite{wineland:blatt,monroe:2007} and
with neutral atoms \cite{Yavuz:2006}. The realization of a
two-qubit gate can be based on collisions between bosonic atoms
\cite{zoller:1999}. This scheme was realized experimentally with
atoms in an optical lattice \cite{bloch:2003}. Fermionic atoms in
the same internal state cannot collide due to the Pauli exclusion
principle. However, atoms in different states can have a large
collisional shift, which can be used to engineer two-qubit SWAP
(or $\sqrt{\text{SWAP}}$) gates, as proposed in ref.
\cite{Hayes:2007}. The scattering length can be made very large by
tuning closer to the Feshbach resonance. For a set of parameters
$\omega =2\pi \times 1$ kHz and $a_{s}\sim 330a_{0}$, a high
fidelity $\sqrt{\text{SWAP}}$ gate can be implemented in a time
period of about $40$ ms. We envision that with an array of many
qubits, a sequence of $\sqrt{\text{SWAP}}$ operations can build
scalable entanglement in this system. This two-qubit operation,
together with single-qubit rotations, provides a set of universal
quantum gates.

The detection of each qubit state at the end of a quantum
computation can be accomplished by spatially-resolved fluorescent
imaging \cite{Meschede:2009,weiss:2007}. This stage must also
resolve the spin of each location. This can be accomplished with
the same method that was used to separate the spin pair, followed
by fluorescent imaging.

One important consideration is the limitation on fidelity set by
decoherence.  One fundamental source is light scattering by the
optical tweezers.  The scattering rate is proportional to the
third power of the frequency.  This was the motivation for using a
$\text{CO}_2$ laser for optical trapping \cite{thomas:2000}. In
our case, such a laser can be used for trapping in two transverse
directions where the trap depth must be higher.  This would
confine the atoms along a line.  However, the optical tweezers
that implement logic gates must be at a shorter wavelength in
order to have sufficient spatial resolution.  The trap depth in
this direction need only be enough to contain the ground state, so
we estimate that light scattering will not limit the quoted
fidelity.  Other sources of decoherence are laser intensity noise
and pointing noise.  The fundamental limit in this case is shot
noise, and should also not limit the predicted fidelity.

In conclusion, we find that fermionic atoms can be used to realize
ultra-high fidelity for quantum computing.  Future work will
analyze the optimum strategies for pairwise entanglement  in this
system.

\begin{acknowledgments}
We thank Jungsang Kim for helpful discussions about MEMS
technology and thank Rob Clark for insightful comments on the
manuscript. M.G.R. was supported by the NSF, the R.A. Welch
Foundation, and the Sid W. Richardson Foundation. C.Z. was
supported by WSU Startup fund and the ARO. Q.N. was supported by
the NSF, the DOE, and the R.A. Welch Foundation.
\end{acknowledgments}

\pagebreak

\pagebreak

\end{document}